\documentclass[prd,aps,twocolumn,floatfix,nofootinbib]{revtex4-1}
\usepackage{graphicx}
\usepackage{xspace}
\usepackage{amssymb}
\usepackage{amsmath}
\usepackage{natbib}
\usepackage[colorlinks]{hyperref}
\def\gr{$\gamma$-ray}
\def\fermi{\textit{Fermi}\xspace}
\def\flat{\textit{Fermi}/LAT\xspace}
\def\ag{\textit{e-ASTROGAM}\xspace}

\def\vek#1{\left(\vcenter{\halign{\hfil$##$\hfil\cr \spvecA#1;;}}\right)}
\def\spvecA#1;{\if;#1;\else #1\cr \expandafter \spvecA \fi}

\begin{document}

\title{Improved limit on axion-like particles from \gr\ data on Perseus cluster}
\author{D. Malyshev$^1$, A. Neronov$^2$, D. Semikoz$^{3,4}$, A. Santangelo$^1$, J. Jochum$^5$}
\address{
$^{1}$ Institut f{\"u}r Astronomie und Astrophysik T{\"u}bingen, Universit{\"a}t T{\"u}bingen, Sand 1, D-72076 T{\"u}bingen, Germany\\
$^2$ISDC, Astronomy Department, University of Geneva, Ch. d'Ecogia 16, 1290, Versoix, Switzerland\\
$^3$ APC, Universite Paris Diderot, CNRS/IN2P3, CEA/IRFU, Observatoire de Paris, Sorbonne Paris Cite, Paris, France\\
$^4$National Research Nuclear University MEPHI (Moscow Engineering Physics Institute),  Moscow, Russia \\
$^5$Physikalisches Institut, Eberhard Karls Universit\"at T\"ubingen, T\"ubingen, Germany}
\begin{abstract}
Coupling of axion-like particles (ALPs) to photons in the presence of background magnetic field affects propagation of \gr s through magnetized environments. This results in modification in the \gr\ spectra of sources in or behind galaxy clusters. We search for the ALP induced effects in the \flat and MAGIC telescope spectra of the radio galaxy NGC 1275 embedded in Perseus galaxy cluster. We report an order-of-magnitude improved upper limit on the ALP-photon coupling constant in the 0.1-10 neV mass range from non-detection of the ALP imprints on the \gr\ spectra. The improved upper limit extends into the coupling range in which the ALP particles could form the dark matter. We estimate the sensitivity improvements for the ALP search achievable with extension of the measurements to lower and higher energies with e-ASTROGAM and CTA and show that the \gr\ probe of ALPs with masses in $10^{-11}-10^{-7}$~eV range will be have order-of-magnitude better sensitivity compared to ground-based experiment IAXO.  
\end{abstract}

\maketitle
\section{Introduction}
\label{sec:intro}

Axions or, more generally, Axion-Like Particles (ALPs) are light weight particles appearing in a range of quantum field theory models as pseudo-Nambu-Goldstone bosons associated to the flat directions in spontaneous breaking of approximate symmetries.  Axions were first introduced in the context of QCD strong CP problem \cite{peccei,weinberg78,wilczek78}. The QCD axions are characterized by a specific relation between mass $m_a$ and photon-coupling constant $g$. More general examples for which the mass and coupling constant are not explicitly related to each other arise in a range of beyond the Standard Model theories (see e.g.\cite{marsh17} for a recent review). Being associated to spontaneous symmetry breaking, ALPs could have been produced in the early Universe via "misalignment" mechanism at the epoch of the symmetry breaking. Such cosmologically produced ALPs could provide sizable contribution to the dark matter (DM, see e.g. ~\citep{preskill83,abbott83} and \cite{duffy09}). A generic necessary condition for the misalignment mechanisms to yield sufficient amount of DM imposes a mass-dependent limit
on the axion coupling to matter \cite{ringwald} 
\begin{equation}
g<10^{-12}\left[\frac{m_a}{1\mbox{ neV}}\right]^{1/2}\mbox{GeV}^{-1}.
\end{equation}

ALPs can be detected in astrophysical observations or laboratory experiments via radiative decays and photon-ALP oscillations in the presence of an (electro)magnetic field (Primakoff process~\cite{primakoff}; see also~\cite{raffelt96}). The non-detection of two photons ALP decay line from DM-dominated objects was used to constraint $g$ for ALPs with masses in the eV-keV range~\cite{ressell91,ringwald14}. ALP-photon oscillations in the presence of magnetic fields are used in  direct-search experiments~\cite{alps2_ex,cast_ex,admx_ex,iaxo_ex,madmax_ex}. 

The same oscillation phenomenon  is also expected to produce features in spectra of astrophysical objects. More specifically, photon-to-ALP conversion is expected to lead to a detectable energy-dependent distortion of the \gr\ spectra of sources in or behind galaxy clusters \cite{ngc1275_xray_limits,fermi_ngc1275}. Intermittent photon-ALP-photon conversion could produce an observable effect of longer-than-expected propagation distance of very-high-energy $\gamma$-rays coming from distant TeV sources \cite{hess_pks2155,axion_cta}. Non-observation of ALP imprints on photon signals from astronomical sources has been used to derive limits on the axion coupling strength (see e.g.~\cite{ngc1275_xray_limits,m87_xray_limits,fermi_ngc1275,hess_pks2155,lower_limits,sn1987_limits} and \cite{chelouche08}).

In this paper we focus on the search of ALP imprints on the \gr\ spectrum of the nearby bright active galactic nucleus NGC 1275, which is embedded in the Perseus galaxy cluster. Two \gr\ sources are found in this nearby cluster: the central radio galaxy NGC 1275 \cite{fermi_ngc1275,magic_ngc1275} and head-tail radio galaxy IC 310 \cite{ic310,IC310_MAGIC}. We consider only the signal  from NGC 1275 because of its higher statistics which leads to a better sensitivity to ALP effects. 

Search for the imprint of the photon-ALP conversion on the \gr\ spectrum of NGC 1275 has been previously reported by \citet{fermi_ngc1275} based on the analysis of the data of Fermi Large Area Telescope (\flat). Our analysis improves significantly the results of  \citet{fermi_ngc1275}, since extends the analyzed energy range toward higher energies, using both the \flat data and the data of ground-based \gr\ telescope MAGIC \cite{magic_ngc1275}. As it will be shown below the combination of \flat and MAGIC spectral measurements provides a possibility to search not only for irregularities of the \gr\ spectrum, induced by the photon-ALP oscillations, but also for the average step-like  suppression of the source flux at high energies. 

In what follows we briefly summarize in Section \ref{sec:data_modelling} the  effects produced by photon-ALP oscillations in the \gr\ spectra. In Section~\ref{sec:data_analysis} we describe the combined \flat and MAGIC spectrum and explain the method used for the modeling of the ALP effect on the \gr\ spectrum.  We discuss the improved constraints on the ALP-photon coupling in the mass range between 0.1 and 10 neV in Section \ref{sec:discussion}.  

\section{Gamma-ray-ALP conversion in galaxy cluster environment}
\label{sec:data_modelling}

The \gr-ALP  conversion probability $P_{\gamma\rightarrow a}$ is a function of axion mass $m_a$, coupling constant $g$ and photon energy $E$. It also depends on the properties of magnetic field $B$ through which \gr s propagate. 

The oscillations of the photon $(A_x, A_y)$ and ALP $a$ fields propagating along the $z$ direction are described by the equation~\citep{hochmuth07,mirizzi09} 
\begin{align}
\label{eq:propagation}
& ( E - i\partial_z - M )\vec{A} = 0 \\ \nonumber
& \vec{A} = \vek{A_x ;A_y ; a}
\end{align}
where $M$ is the mixing matrix 
\begin{align}
\label{eq:mixing_matrix}
& M = \begin{bmatrix}
\Delta_{11} & \Delta_{12} & \Delta_{a\gamma}c_\phi \\
\Delta_{12} & \Delta_{22} & \Delta_{a\gamma}s_\phi \\
\Delta_{a\gamma}c_\phi & \Delta_{a\gamma}s_\phi & \Delta_a
\end{bmatrix} 
\end{align}
with
\begin{align}
& \Delta_{11} = \Delta_{\parallel}c_\phi^{2} + \Delta_{\perp}s_\phi^{2}  \nonumber\\
& \Delta_{22} = \Delta_{\parallel}s_\phi^{2} + \Delta_{\perp}c_\phi^{2} \nonumber\\
& \Delta_{12} = (\Delta_{\parallel} - \Delta_{\perp})s_\phi c_\phi \nonumber\\
& \qquad\qquad  \nonumber\\
& \Delta_{\parallel} = \Delta_{pl} + \frac{7}{2}\Delta_{QED} \nonumber\\
& \Delta_{\perp} = \Delta_{pl} + 2\Delta_{QED} \nonumber 
\end{align}
The main contributions to the mixing matrix for the range of parameters relevant to GeV-TeV photons propagating through galaxy cluster environment come from
\begin{align}
& \Delta_{a\gamma} = \frac{gB_T}{2}\approx \frac{1}{70\mbox{ kpc}} \left[\frac{g}{10^{-12}\mbox{ GeV}^{-1}}\right]\left[\frac{B_T}{10\ \mu\mbox{G}}\right]  \nonumber\\
& \Delta_a = -\frac{m_a^{2}}{2E}\approx-\frac{1}{133\mbox{ kpc}}\left[\frac{m}{1\mbox{ neV}}\right]^2\left[\frac{E}{10\mbox{ GeV}}\right]^{-1} \nonumber
\end{align}
while $\Delta_{pl}$ and $\Delta_{QED}$ are small and could be neglected considering the GeV-TeV \gr\ propagation through the galaxy cluster environment. Only the magnetic field component $B_T$ transversal to the direction of photon/ALP propagation is relevant. The transverse magnetic field $B_T$ orientation in the $xy$ plane is described by the cosine $c_\phi$ and sine $s_\phi$ of the field inclination with respect to the $x$ axis. 

If the magnetic field and plasma density do not vary in space, Eq.~\ref{eq:propagation} could be solved analytically. In this case the expression for  the probability of photon to ALP conversion is~\citep{hochmuth07}:
\begin{eqnarray}
\label{eq:constB_propagation}
&& P_{\gamma\rightarrow a}(s)\simeq (\Delta_{a\gamma} s)^2 \frac{\sin^2\Delta_{osc}s}{(\Delta_{osc}s)^2} \\ \nonumber
&& \Delta_{osc}^2 \simeq \Delta_{a}^2 + 4\Delta_{a\gamma}^2
\end{eqnarray}
where $s$ is the propagation distance. The two contributions to $\Delta_{osc}$ become equal at the energy
\begin{equation}
\label{eq:ecr}
E_{cr}\simeq 2.5\left[\frac{m_a}{1\mbox{ neV}}\right]^2\left[\frac{g}{10^{-12}\mbox{ GeV}^{-1}}\right]^{-1}\left[\frac{B_T}{10\ \mu\mbox{G}}\right]^{-1}\mbox{ GeV}
\end{equation}

If the magnetic field varies on the distance scale $\delta z$ and the photon-ALP conversion probability $P_0$ on this distance scale is $P_0\ll 1$, the photon-ALP conversion probability after the passage of the primary \gr\ beam through $N$ domains of  size $\delta z$ is~\citep{hochmuth07}
\begin{equation}
\label{eq:analytic}
P_{\gamma\rightarrow a} \simeq \frac{1}{3}\left(1-\exp(-3NP_0/2)\right)
\end{equation}
It saturates at $P_{\gamma\rightarrow a}\simeq 1/3$ in the limit $NP_0\gg 1$.

The conditions $P_0\ll 1$ and/or $P_0N\gg 1$ are not necessarily met for the relevant range of photon energies $E$ and ALP parameters $g,m_a$. In this case Eq.~\ref{eq:propagation} has to be solved numerically. 

For this numerical solution, we consider $N_{sim}=1000$ realizations of the cluster  magnetic field which randomly changes direction on the distance scale $\delta z=10$~kpc. Following \citet{fermi_ngc1275} we choose the magnetic field strength in the cluster core to be $B_0\simeq 15\ \mu$G and assume the strength of the field to vary with together with the electron density $B\propto n_e^{0.5}$. The electron density profile corresponds to one, derived from X-ray observations \citep{churazov}. This results in the field strength dependence on the distance from the cluster center 
\begin{align}
\label{eq:field_profile}
&B(r) = \begin{cases}B_0 & if\quad r\leq r_0 \\
B_0\cdot(r/r_0)^{-0.5} & if\quad r>r_0
\end{cases} \\ \nonumber
&B_0=15\mu G; \quad r_0=40\mbox{~kpc}
\end{align}

For each field realization we consider a range of axion's masses $m_a=10^{-11}..10^{-7}$~eV and coupling coupling constants $g=10^{-14}..10^{-9}$~GeV$^-1$ to find $P_{\gamma\rightarrow a}(E)$. We then model the observed spectrum as 
\begin{align}
\label{eq:spectral_model}
& F(E) = F_0(E)\cdot \left(1 - P_{\gamma\rightarrow a}(E) \right) 
\end{align}
where $F_0(E)$ is the reference spectral model of NGC 1275  \flat $+$ MAGIC time-average spectrum which provide the best description of the data without the account of possible ALP effects. 

\begin{figure*}
\includegraphics[width=1.\linewidth]{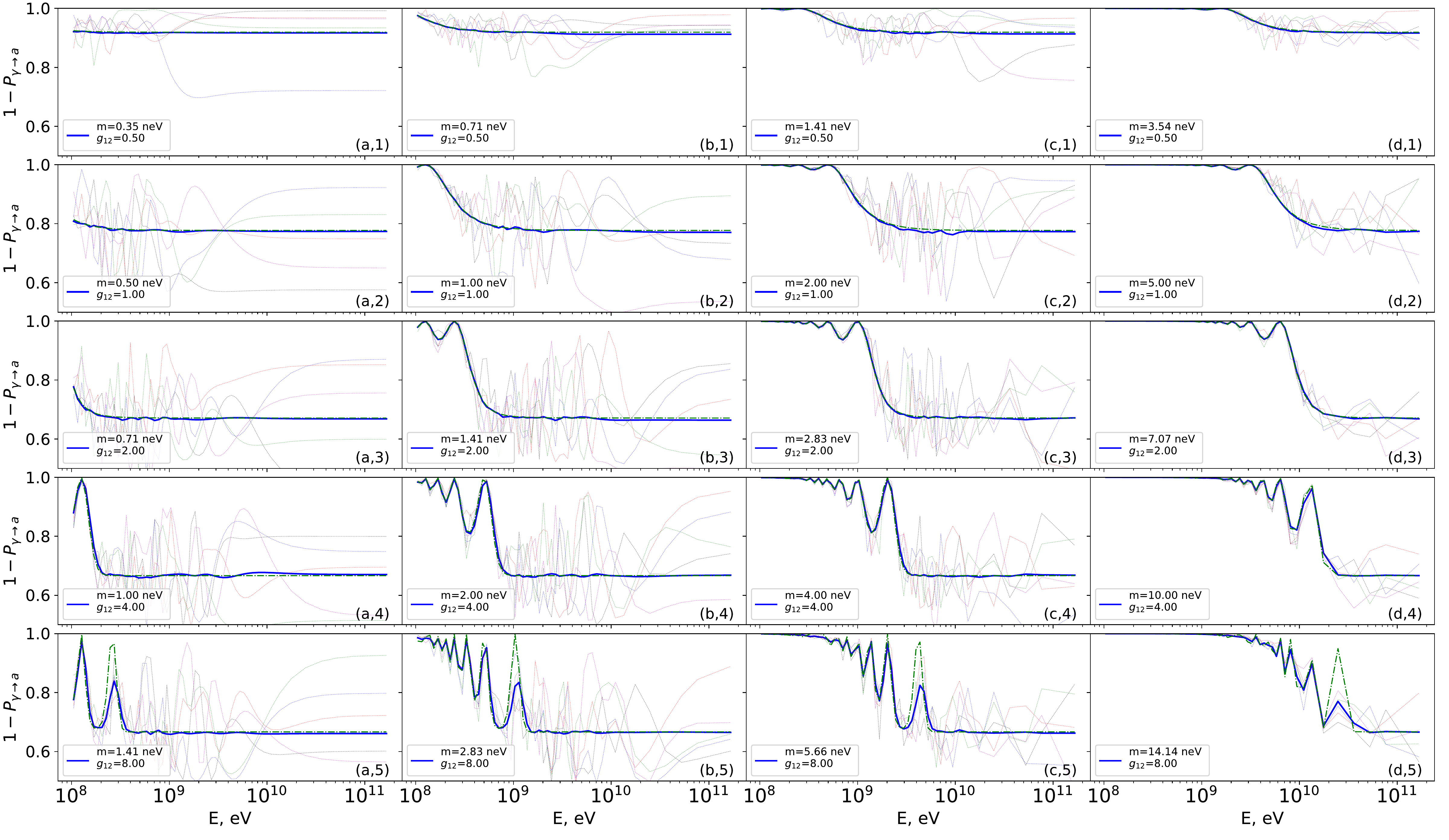}
\caption{Absorption curves ($1-P_{\gamma\rightarrow a}$) for various  $(m_a ; g)$ parameters (indicated in the panel legend and corresponding to the markers in the first panel of Fig. \ref{fig:exclusion_comparison}). Thick solid blue curves in each panel show the value of ($1-P_{\gamma\rightarrow a}$) averaged over different realizations of the magnetic field. Thin curves show the values of ($1-P_{\gamma\rightarrow a}$)  for 5 randomly selected realizations. Dot-dashed green curves illustrate 1-$P_{\gamma\rightarrow a}$ from Eq.~\ref{eq:analytic} for characteristic magnetic field $B=8$~$\mu$G, cluster size $N\delta z=0.5$~Mpc and magnetic filed domain size $\delta z = 10$~kpc. See also Fig.~\ref{fig:exclusion_comparison} for the location of corresponding curves on exclusion plot. }
\label{fig:absorption_curves}
\end{figure*}

Examples of $(1-P_{\gamma\rightarrow a})$ curves for different values of $m_a$ and $g$ are shown in Fig. \ref{fig:absorption_curves}. One could see that the effect of photon-ALP conversion on the \gr\ signal has two characteristic signatures. First, the conversion introduces an overall flux suppression in the energy range $E>E_{cr}$. This suppression is down to the factor $2/3\simeq 0.66$ for high values of $g$, but it becomes weaker for small values of $g$. In this regime $P_0\ll 1$ and $\Delta_{osc}\delta z\ll 1$. In this case 
\begin{equation}
P_0\simeq (\Delta_{a\gamma}\delta z)^2\simeq \frac{g^2B_T^2(\delta z)^2}{4}
\end{equation}
The argument of the exponent in Eq. (\ref{eq:analytic}) becomes  small $NP_0\ll 1$ and an approximate expression for $P_{\gamma\rightarrow a}$ is
\begin{eqnarray}
&&P_{\gamma\rightarrow a}\simeq \frac{Ng^2B_T^2(\delta z)^2}{8}\\ &&\simeq 0.1\left[\frac{N\delta z}{100\mbox{ kpc}}\right]\left[\frac{\delta z}{10\mbox{ kpc}}\right]\left[\frac{g}{10^{-12}\mbox{ GeV}^{-1}}\right]^2\left[\frac{B_T}{10\ \mu\mbox{G}}\right]^2\nonumber
\end{eqnarray}
If the quality of the spectral measurements in the GeV-TeV band is at the 10\% level, the 10\% step-like flux suppression of the type shown in Fig. \ref{fig:absorption_curves} could be measured and the ALP coupling values down to $g\sim 10^{-12}$~GeV$^{-1}$ could be probed, provides that the intrinsic source spectrum is smooth enough to allow the detection of step-like features.  

The range of the ALP masses which could be probed via the search of the step-like spectral features is limited by the condition that the critical energy $E_{cr}$ should be well within the energy range of the measurements, which is $100$~MeV$\ll E_{cr}\ll 1$~TeV in our analysis. This defines the mass range $0.1$~neV$\ll m_a\ll 10$~neV (see Eq. \ref{eq:ecr}).

Another type of spectral distortions introduced by photon-ALP conversion is the oscillatory behavior of the spectrum. Such behavior is generically present in the energy range around $E_{cr}$. Search for such oscillatory behavior was the focus of the analysis of \citet{fermi_ngc1275}. From Fig. \ref{fig:absorption_curves} one could see that the strength of the oscillatory features diminishes with the decrease of $g$. The amplitude of the oscillations drops below $\sim 10\%$ around the average curve for $g\lesssim 10^{-11}$~GeV$^{-1}$. This range of $g$ corresponds to the sensitivity limit of the analysis reported by \citet{fermi_ngc1275}. 

Our analysis described below focuses on the search of both oscillatory and step-wise modifications of the spectrum generated by  photon-ALP conversion. Combining the two signatures we can reach higher sensitivity, down to $g\sim 10^{-12}$~GeV$^{-1}$. 

\section{Data analysis}
\label{sec:data_analysis}
\subsection{\flat data analysis}

\flat data selected for the analysis presented in this paper cover more than 9 years (Aug. 2008 to Sept. 2017). We use the latest available \fermi Science Tools \texttt{v10r0p5} with P8\_R2 response functions (\texttt{CLEAN} photon class)\footnote{See \href{https://fermi.gsfc.nasa.gov/ssc/data/analysis/documentation/Cicerone/Cicerone_LAT_IRFs/c }{description of \flat response functions} }. 

The spectral features introduced by photon-ALP conversion appear as deviations of the measured spectrum from the smooth broadband intrinsic source spectrum, which is typically described by a powerlaw with only broad band features (breaks, cut-offs) reflecting the specificity of the physical processes inside the source. The highest sensitivity of the ALP search is achieved if the intrinsic source spectrum is a powerlaw without additional intrinsic spectral features which can mimic some of photon-ALP signatures. The slope of the powerlaw \gr\ spectra of AGN is known to vary with the flux, with faster spectral variability occurring during flaring periods. Such flare-related spectral variability might introduce spectral features in the time-average spectra. This might result in a reduced sensitivity for ALP search. Taking this into account, we performed timing analysis  excluding flaring time intervals from the dataset. We extracted lightcurves of the sources in the 0.1 -- 300~GeV energy band and remove time bins with flux exceeding the mean level at more than $2.5\sigma$.

We performed the standard binned likelihood analysis of a region around the source to extract the time-averaged spectrum. The spectral analysis is based on the fitting of the spatial / spectral model of the sky region around the source of interest to the data. The region-of-interest considered in the analysis is a circle of radius 18 degrees around the Perseus cluster.  The model of the region included all sources from the 3FGL catalogue as well as components for isotropic and galactic diffuse emissions given by the standard spatial/spectral templates \texttt{iso\_P8R2\_CLEAN\_V6\_v06.txt} and \texttt{gll\_iem\_v06.fits}. The spectral template for each 3FGL source in the region was selected according to the catalogue model. The normalizations of the sources were considered to be free parameters during the fitting procedure. Following the recommendation of the \flat collaboration, we performed our analysis enabling energy dispersion handling.

We performed the spectral analysis in a set of narrow energy bins. In order to have significant photon statistics in each of these bins as well as keep their width not to exceed \flat energy resolution we defined an adaptive energy binning using the following prescription. We required that each energy bin had to contain at least 100~photons in a circle of 1~deg radius around the source and the bin upper energy bound had to be at least 10\% of its lower energy bound. Such a choice resulted in the energy binning following the instrument energy resolution up to   $\lesssim 10$~GeV and in a coarser binning at higher energies. 
Our analysis includes statistical errors as well as systematic uncertainty at 3\% flux level for energies $\leq 10$~GeV and 5\% above this energy\footnote{See \href{https://fermi.gsfc.nasa.gov/ssc/data/analysis/scitools/ Aeff\_Systematics.html}{description of \flat systematics}}.

\subsection{Broad band spectrum of NGC 1275}
\begin{figure}
\includegraphics[width=\linewidth]{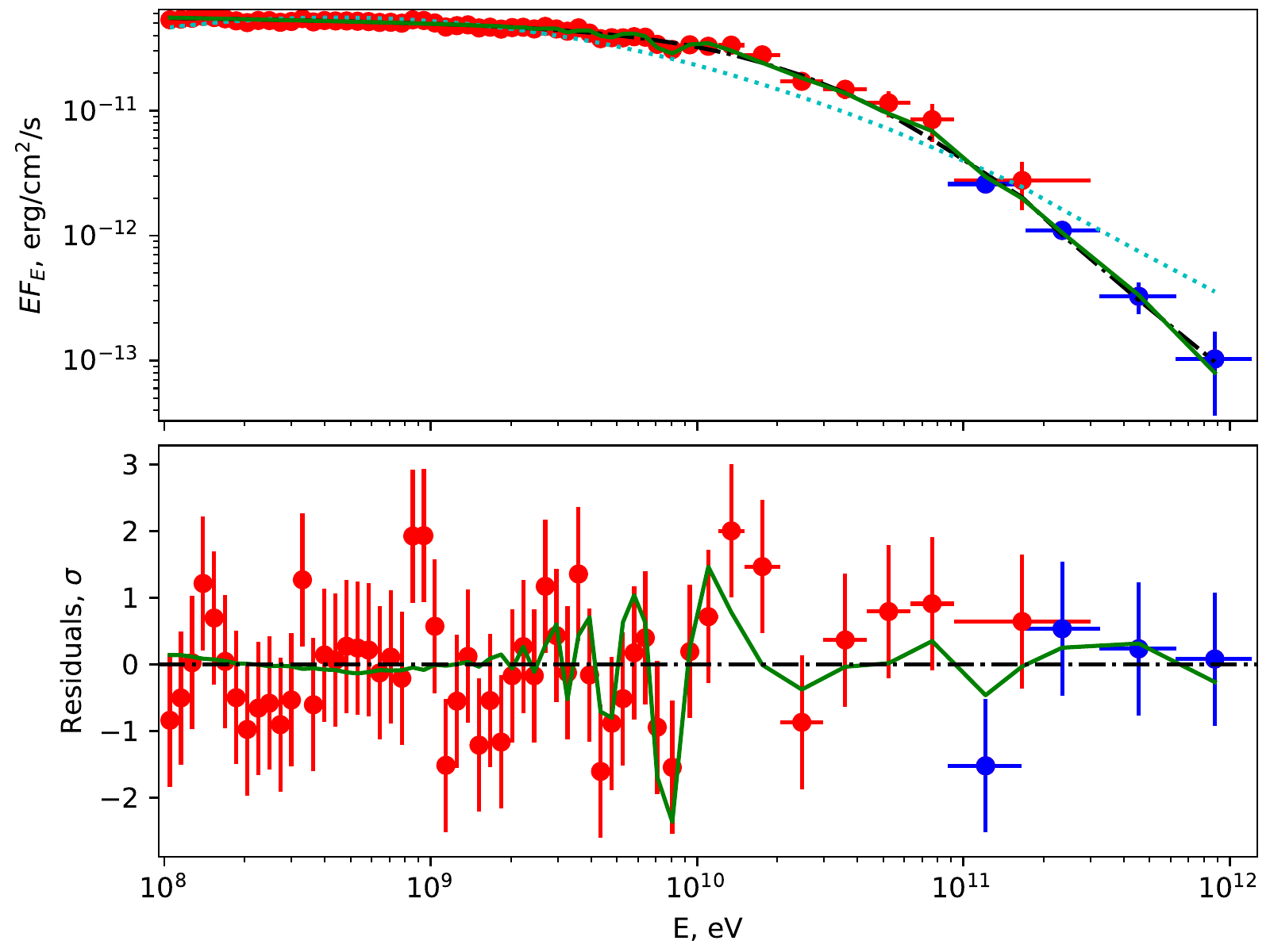}
\caption{Combined \flat(red) and MAGIC (blue) spectrum of NGC 1275 together with the best-fit broken powerlaw model (black dash-dotted line). Green curve shows the model with account of the effect of photon-ALP conversion for $m_a=9.6\cdot 10^{-9}$~eV, $g=4\cdot 10^{-12}$~GeV$^{-1}$. Bottom panel shows the fit residuals with respect to the broken powerlaw  spectral model and the difference between the broken powerlaw model and the best-fit model which includes photon-ALP conversion(green curve). Cyan dashed line illustrates the best-fit log-parabola model. Note, that only statistical errors are shown.}
\label{fig:spectrum_ngc1275}
\end{figure}
The combined \flat and MAGIC spectrum of NGC 1275 is shown in Fig. \ref{fig:spectrum_ngc1275}. MAGIC observation campaign reported by \citet{magic_ngc1275} included long-term exposure of the source over several years, so that the spectral measurements represent, similarly to the \flat measurements, the time-averaged source spectrum. One could see that the two spectra agree well in the overlapping energy range. 

The spectral measurements extend over four decades in energy. We have verified that the log-parabola model of the broadband spectrum considered by \citet{fermi_ngc1275} could not fit the data over such broad energy range (reduced $\chi^2$ of the fit is $\sim 157$ ($\sim 355$ without account for systematics) for 56 degrees of freedom). We therefore considered an alternative broken powerlaw  model
\begin{align}
& F_0(E) \propto \frac{E^\alpha}{\left[1+(E/E_{br})^\kappa\right]^{(\beta+\alpha)/\kappa}} 
\end{align}
We find that the best-fit is provided by the broken powerlaw model with break energy $E_{br}=35\pm 15 $~GeV, the low-energy slope $\alpha=-2.04\pm 0.02$, high-energy slope $\beta=3.9\pm 0.4$ and  $\kappa = 1.1\pm 0.2$. The best-fit $\chi^2$ value is  $\chi^2_0=19.04$ for 54 degrees of freedom ($\chi^2_0=44.46$ without account for systematics). We use the broken powerlaw model as a reference model for the search of the ALP-induced spectral features.

As it is discussed in the previous section, there are two types of the spectral features induced by photon-ALP conversion: the oscillatory behavior of the spectrum and the overall suppression of the flux above $E_{cr}$. The presence of the break in the broadband spectrum of NGC 1275 inevitably limits the sensitivity of the search for the step-like suppression of the flux in the energy range $E_{cr}\sim E_{br}$. A step-like flux change above $E_{cr}$ could be confused with a slight shift of $E_{br}$ in this case.

\subsection{Constraints on $m_a,\ g$}
\label{sec:results}

\begin{figure}[t]
\includegraphics[width=\linewidth]{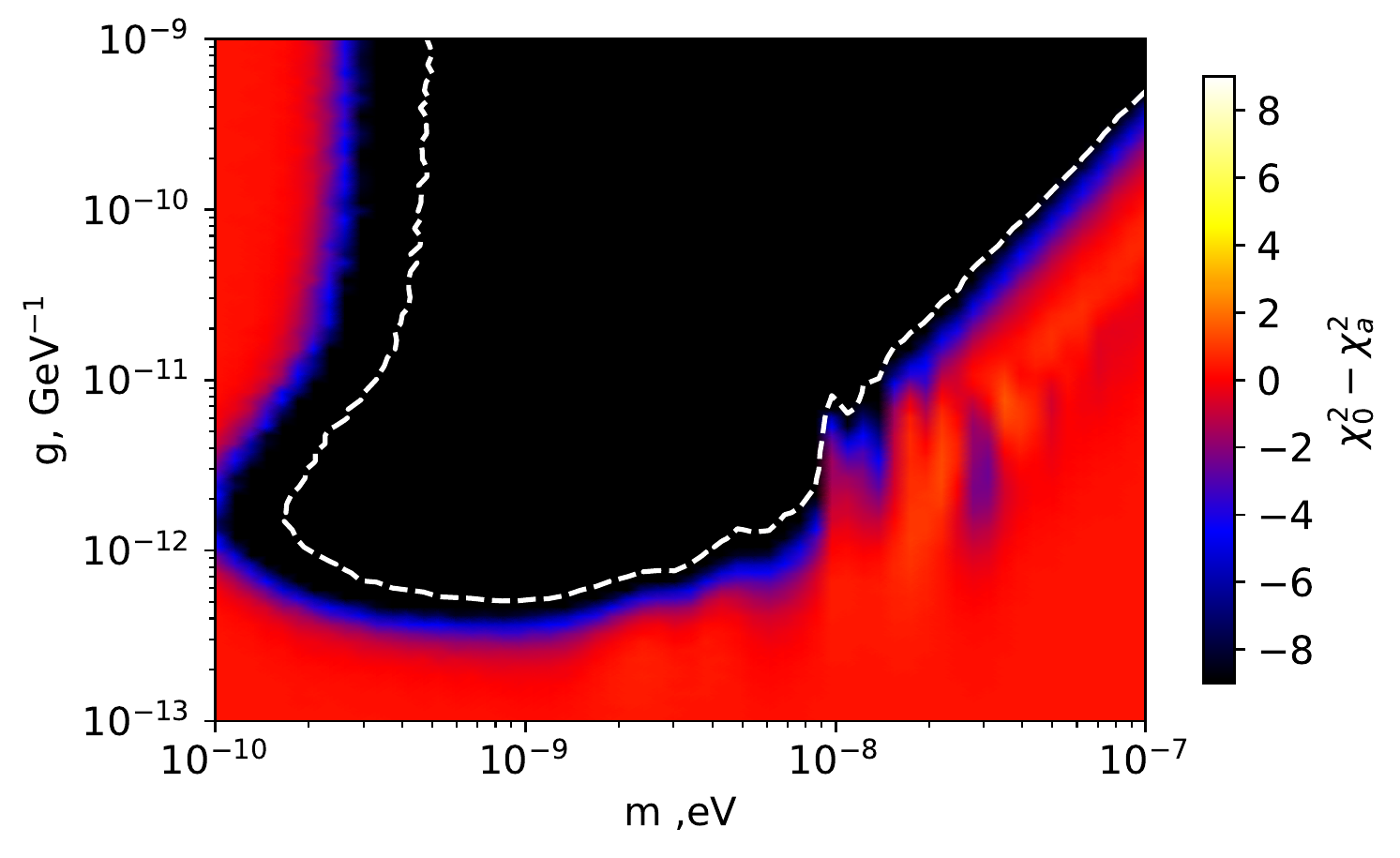}
\caption{The map of median values of $\chi^2_0-\chi^2_a$ distribution (over random realizations of magnetic field in the cluster) for different $m_a,\ g$. The range of parameters for which the presence of the photon-ALP conversion effect is inconsistent with the data  is delimited by white dashed  line. }
\label{fig:exclusion_pers}
\end{figure}

To search for the ALP-induced spectral features, we re-fit the spectrum with model (see Eq.~\ref{eq:spectral_model}) using
 the function $P_{\gamma\rightarrow a}(E)$ calculated for different realizations of the magnetic field and different values of the axion mass and coupling constant. As a result, for each  pair $(m_a ; g_a)$ we obtain a distribution of $N_{sim}=1000$ best-fit $\chi_{a}^2 -\chi^2_{0}$ values, where $\chi^2_a$ is the $\chi^2$ value of the fit taking into account photon-ALP conversion function $P_{\gamma\rightarrow a}(E)$. The median value of this distribution is shown for each pair $m_a, g$ in Fig. \ref{fig:exclusion_pers}.

Addition of photon-ALP conversion effect could occasionally improve the $\chi^2$ of the fit, compared to the model without axion. This is illustrated in Fig. \ref{fig:spectrum_ngc1275} where a fit of the spectrum with the photon-ALP conversion effect is shown for $m_a=9.6\cdot 10^{-9}$~eV, $g=4\cdot 10^{-12}$~GeV$^{-1}$. Without account for systematic uncertainty the fit has $\chi^2_a-\chi^2_0=-7.35$ which corresponds to an improvement significant at $\sim 2.2\sigma$ level for the chance coincidence probability for the fit improvement for nested models with two added parameters. 

Alternatively, addition of the photon-ALP conversion effect worsens the fit, because it introduces oscillations  and/or the step-like change in the spectrum which is not observed. In this case, the $\chi^2$ value of the fit grows. An increase of the $\chi^2$ larger than by 6.2 corresponds to $2\sigma$ level inconsistency of the model with the data.  

However, the increase of the $\chi^2$ depends on the magnetic field realization. The inconsistency of the model with the data might be at $>2\sigma$ level for some magnetic field realizations, but smaller for others. Since we are considering a single source, we have to take into account the possibility that the magnetic field configuration along the single line of sight might be peculiar so that the photon-ALP conversion effect on the spectrum is diminished. 

We therefore choose the following criterion of inconsistency of the model with the photon-ALP conversion with the data: the parameter pair  $(m_a ; g_a)$ is excluded if the increase of the $\chi^2$ of the fit is by more than 6.2 in  95\% of the realizations of magnetic field.  The range of $m_a, g$ parameters for which the presence of the photon-ALP conversion effect is inconsistent with the data is delimited by the white dashed line in Fig. \ref{fig:exclusion_pers}.

\begin{figure*}
\includegraphics[width=0.49\linewidth]{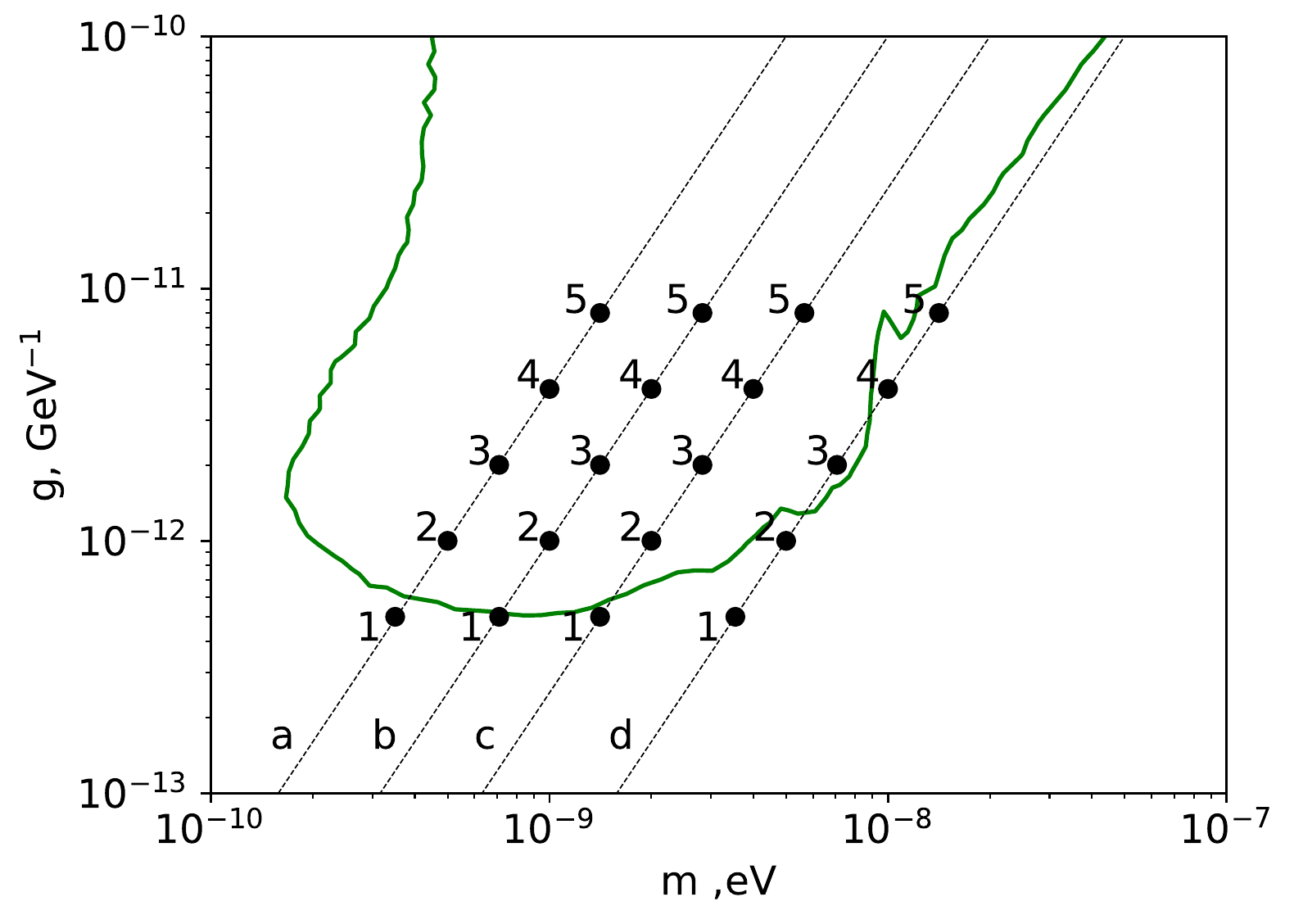}
\includegraphics[width=0.49\linewidth]{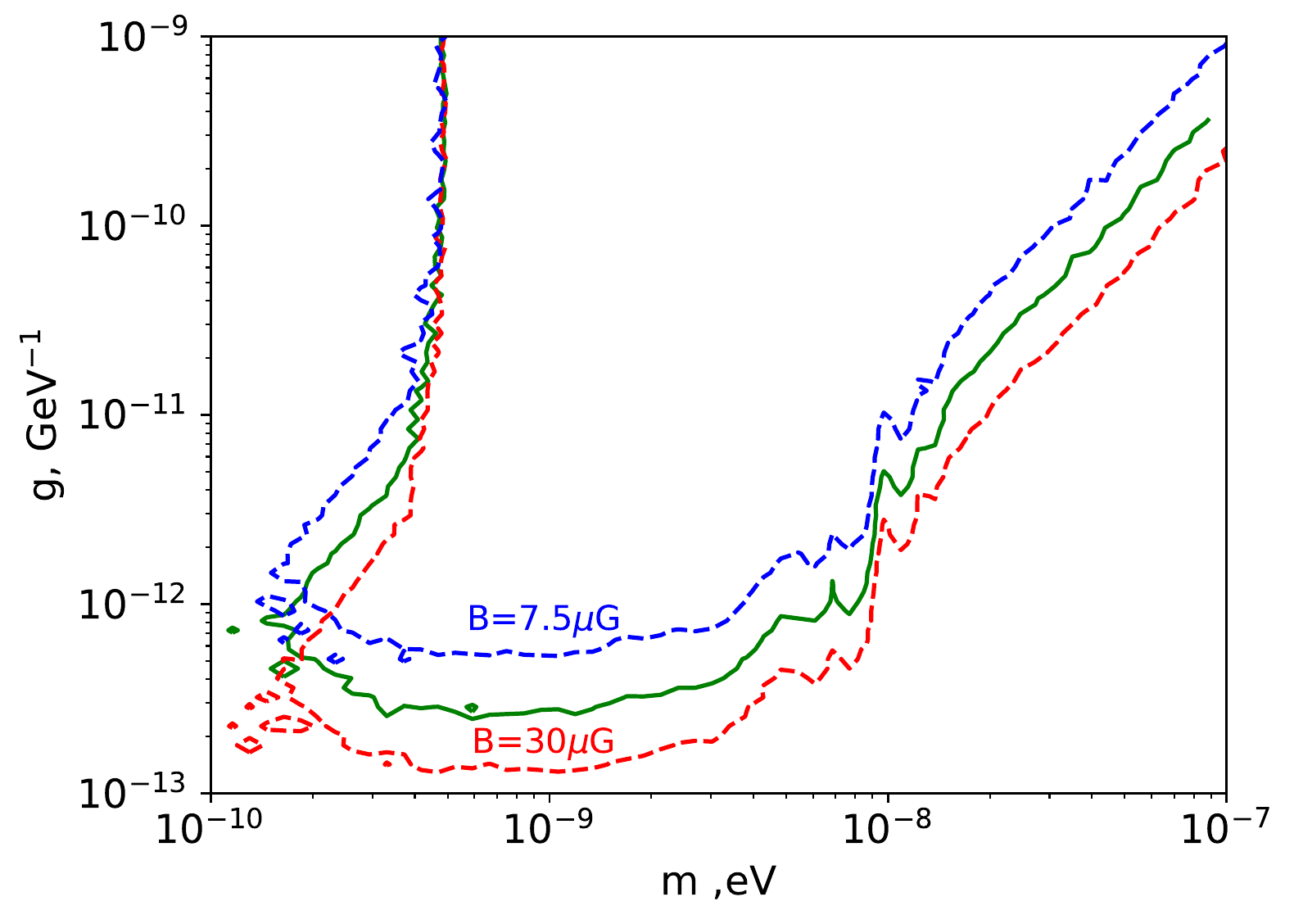}
\includegraphics[width=0.49\linewidth]{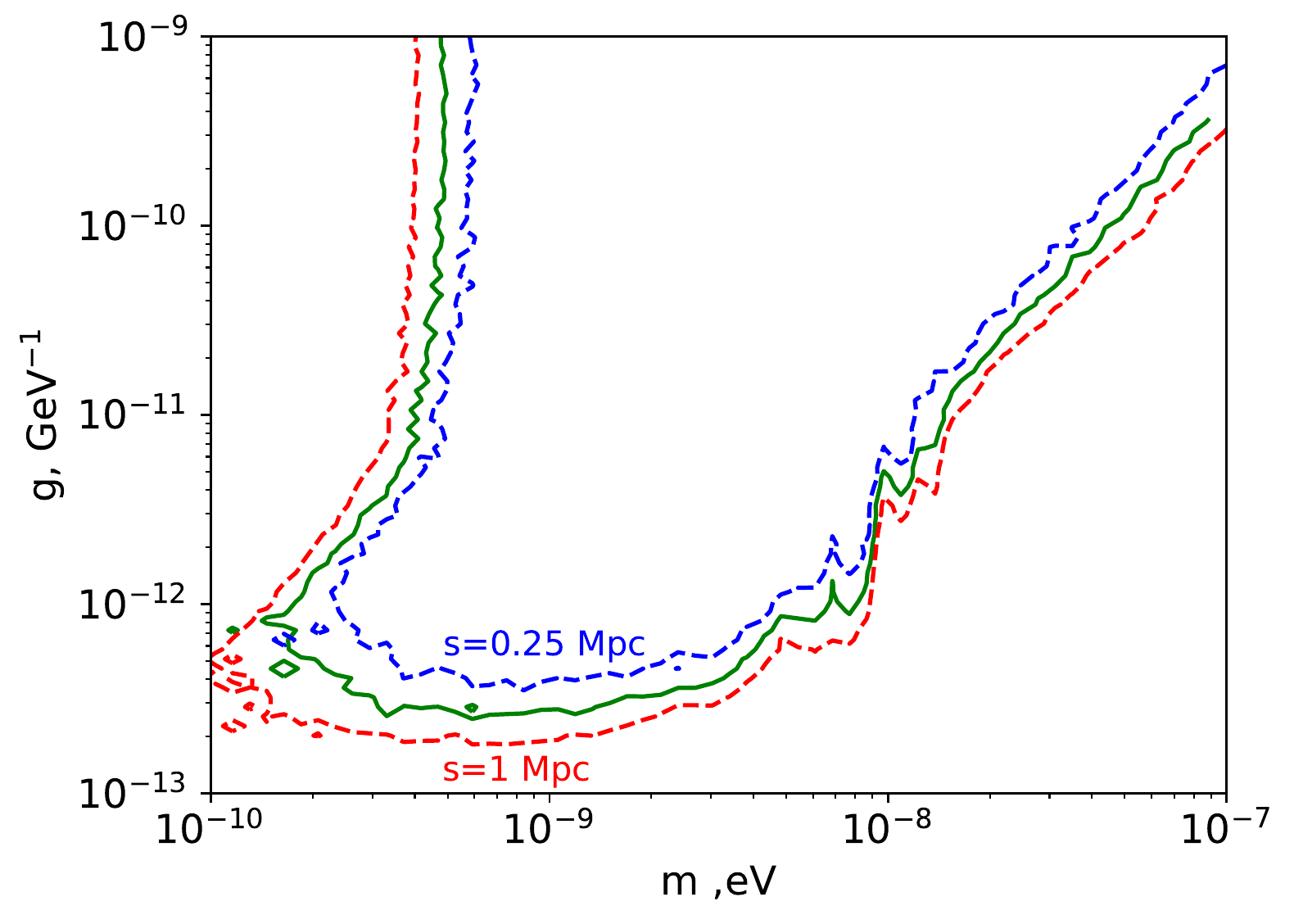}
\includegraphics[width=0.49\linewidth]{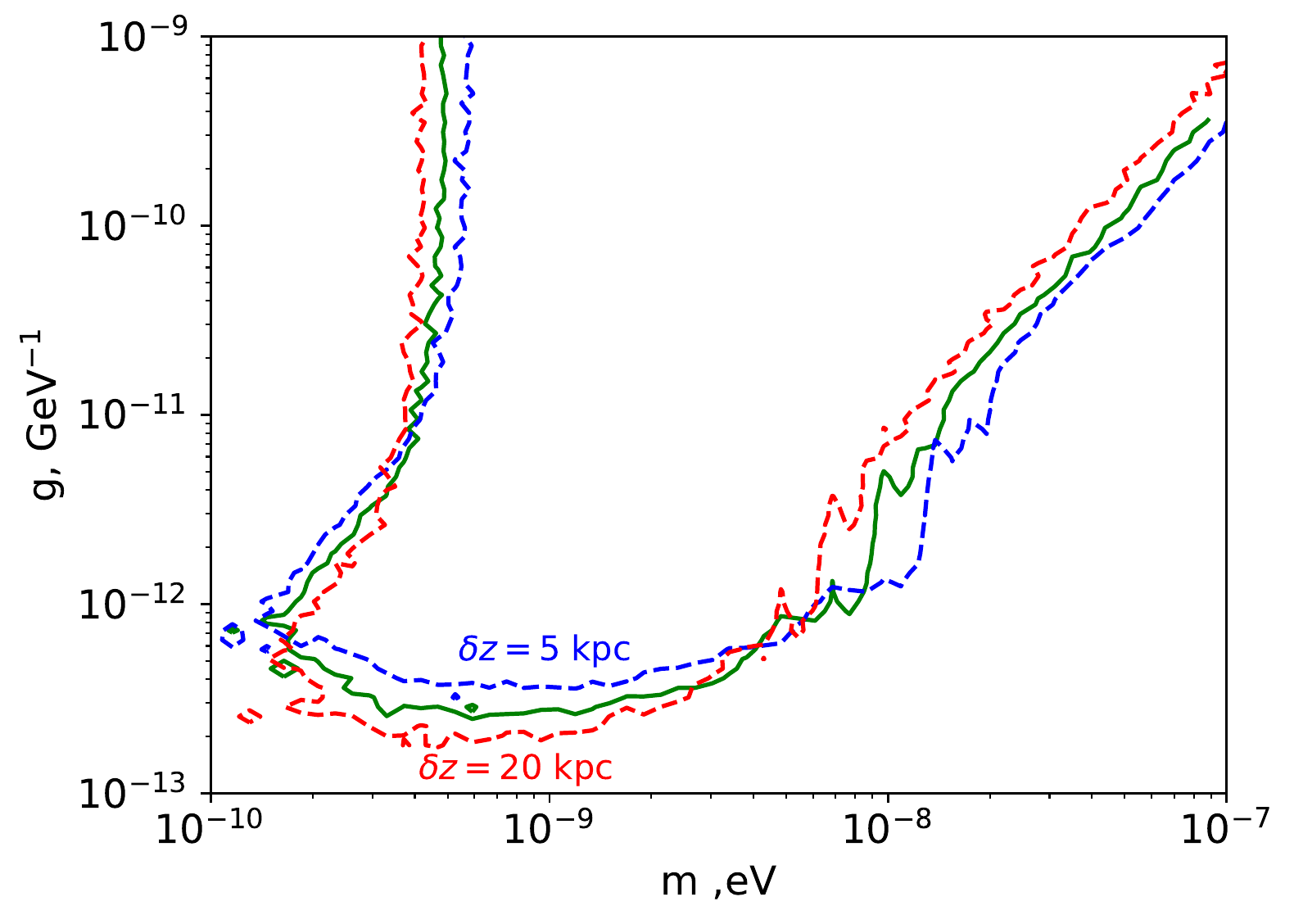}
\caption{Comparison of constraints on $m_a,\ g$ for different parameters of magnetic field. Left top panel shows the constraints obtained for the reference model considered in the text. Dots mark the reference values of $m_a,\ g$ for which the $P_{\gamma\rightarrow a}$ curves are shown in Fig. \ref{fig:absorption_curves}. Inclined lines correspond to $E_{cr}=const$. Top right panel shows modification of the constraints on $m_a, g$ for different values of the magnetic field strength in the cluster core. Bottom left panel shows variations of the constraints for different assumptions on the overall extent of the cluster magnetic field, while the bottom right panel shows the dependence of the constraints on the magnetic field correlation length. }
\label{fig:exclusion_comparison}
\end{figure*}

The extent of the region of $m_a, g$ parameter space excluded by the data depends on the underlying model of magnetic field in Perseus cluster. This model suffers from uncertainties. For example, Faraday rotation data \cite{magnetic_field} indicate that the central magnetic field strength might reach $25\ \mu$G (compared to $B_0=15\ \mu$G with $\sim 10\ \mu$G root mean square assumed in our reference model). The magnetic field correlation length could only be estimated at an order-of-magnitude level from the rotation measure data. The overall extent of the magnetized region is also not well constrained because of the lack of the data at the outskirts of the cluster. Taking this into account, we explore the dependence of the limits on $m_a, g$ on the magnetic field model parameters. The result is shown in Fig. \ref{fig:exclusion_comparison}. One could see that changes in the basic magnetic field model parameters could shift the boundaries of exclusion region by a factor $\simeq 2$.

\section{Discussion}
\label{sec:discussion}
\subsection{Comparison with other limits on $m_a, \ g$}

\begin{figure*}
\includegraphics[width=0.495\linewidth]{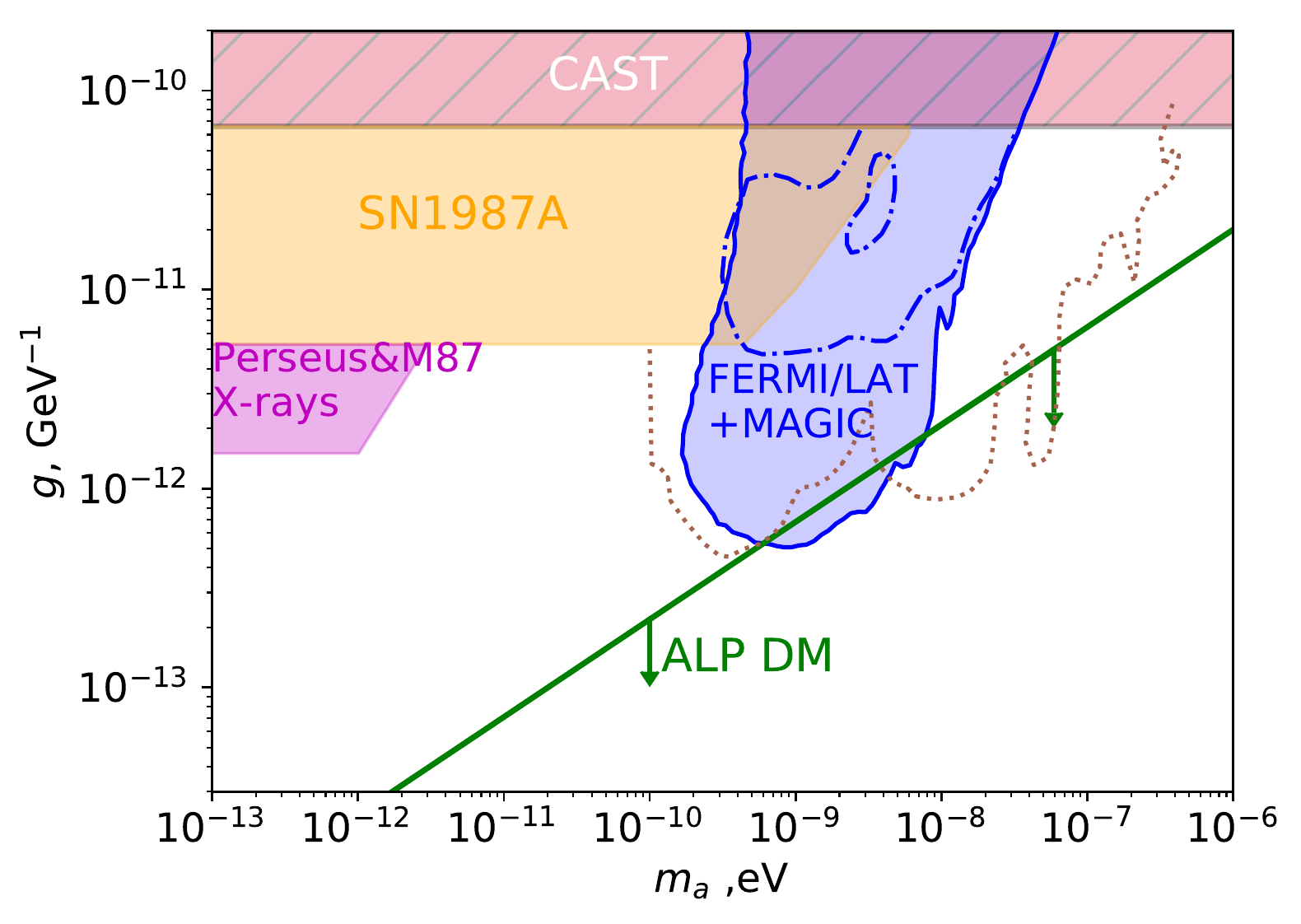}
\includegraphics[width=0.495\linewidth]{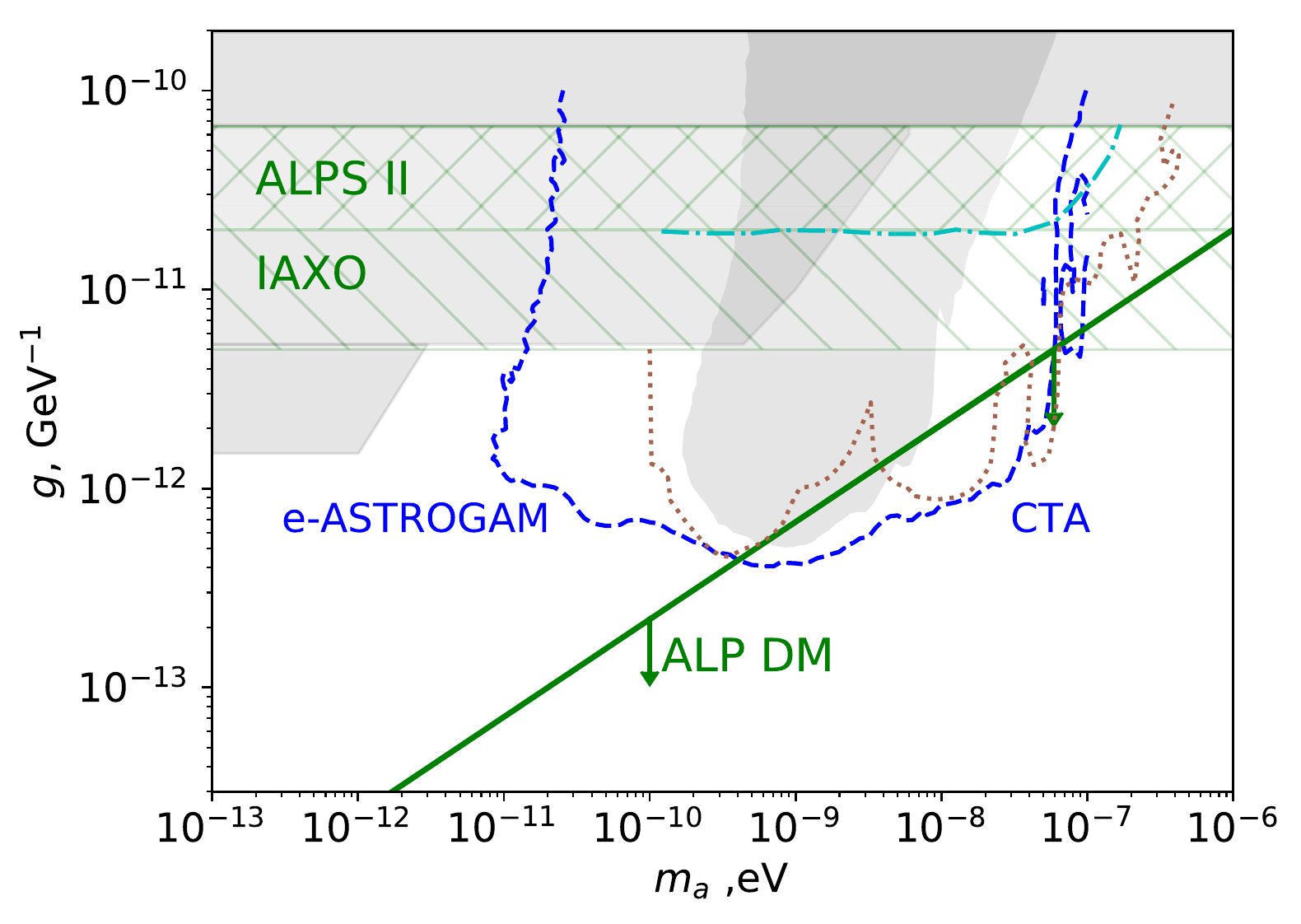}\\
\caption{Comparison of limits on $m_a, g$ from current (left panel) and future (right panel) experiments and observations. \textit{Left:} blue-shaded region shows the constraint derived in the present work.  Horizontal shaded region shows limits on ALP parameters from CAST~\cite{new_cast_limits} experiment. Orange-shaded region shows the limit from non-observation of the ALP effects in SN 1987A~\cite{sn1987_limits}. Magenta shared region shows the constraint from non-observation of the photon-ALP conversion effect in X-rays in the signals of NGC~1275 and M~87~\cite{ngc1275_xray_limits,m87_xray_limits}). Brown dotted curve shows the parameter range in which photon-ALP conversion effects could effectively increase transparency of the Universe to TeV photons~\cite{lower_limits}. Previous \flat exclusion limit~\cite{fermi_ngc1275} is shown by the blue dot-dashed line. \textit{Right:} Grey shaded regions show existing constraints on $m_a, g$ (same as in the left panel). Brown dotted line is the same as in the left panel. Horizontal hatched regions show the sensitivity of future (ALPS~II~\cite{alps2_limits}, IAXO~\cite{iaxo_limits}) experiments. Blue solid curve shows the sensitivity achievable with joint \ag + \flat + CTA observations of Perseus cluster of galaxies. Cyan dash-dotted curve shows the sensitivity of ALP search with CTA based on increased transparency of the Universe for TeV photons~\cite{axion_cta}. 
}
\label{fig:exclusions_all}
\end{figure*}
 
Constraints on the ALP coupling $g$ in the mass range $0.1-10$~neV derived from the combined \flat and MAGIC data on NGC 1275 provide an order-of-magnitude improvement compared to the previous analysis based on \flat data. This improvement is resulted by the extended energy range of the data. This results in a better constrained baseline spectral model without the photon-ALP conversion effect. The better constrained model has allowed the search for the step-like suppression of the source flux above the critical energy, a search method which provides better sensitivity for small values of $g$, as explained in Section \ref{sec:data_modelling}. 

One could also see from Fig. \ref{fig:exclusions_all} that the sensitivity of the \gr\ search of the photon-ALP conversion effect exceeds that of the direct search techniques in the $0.1-10$~neV mass range by two orders of magnitude (compared to CAST)~\cite{new_cast_limits}. In the same figure we show how the measurement reported in this paper compares to the sensitivity of the next-generation direct search experiments ALPS-II and IAXO \cite{alps2_limits,iaxo_limits}.

In this respect, it is interesting to note that the improvement of the sensitivity provided by the combined \flat+MAGIC search  is sufficient to probe the range of $m_a, g$ parameter space in which the misalignment mechanism could in principle result in production of ALP DM \cite{ringwald}. Constraints on ALP dark matter models are shown by the green inclined line in Fig. \ref{fig:exclusions_all}. One could see that the range of parameters excluded by the \flat+MAGIC search reaches the ALP DM region. 

The left panel of the Fig.~\ref{fig:exclusions_all} illustrates also the limits on ALP parameters from several indirect-search approaches (non-observations of ALP effects in 1987A~\cite{sn1987_limits}; similar to presented above analysis of X-ray data of NGC~1275 and M~87~\cite{ngc1275_xray_limits,m87_xray_limits}).

The improved bound on the ALP-photon coupling also excludes significant part of parameter space in which the oscillation effect could result in higher transparency of the Universe for very-high-energy \gr s ~\cite{lower_limits}.

\subsection{Estimate of sensitivity of ALP searches with next-generation $\gamma$-ray telescopes}

The sensitivity reach of the ALP search with \flat and MAGIC is limited by the available energy range. In particular, search for the step-like flux suppression requires that $E_{cr}$ lies well inside the energy range of the measurements. The minimal detectable level of the step-like suppression depends on the statistics of the \gr\ signal. Extension of the energy range and increase of the effective collection area provided by the next-generation telescopes will result in the improvement of the sensitivity of the ALP searches. These improvements are illustrated in Fig. \ref{fig:exclusions_all}.

To estimate the improvement of the sensitivity of the ALP search with next-generation telescopes we consider the \gr\ signal from NGC 1275 detectable with \ag and CTA. 

\ag space-based \gr\ telescope is designed for improvement of sensitivity in the energy range below 100~MeV (compared to \flat). It will be sensitive in the energy range 0.03 -- 3000~MeV~\cite{ag_description,ag_science}. Assuming that the spectrum of NGC 1275 continues as an $E^{-2}$ type powerlaw toward energies $E<100$~MeV, we have simulated  a $30$~Msec observation of NGC 1275 with \ag using publicly available \ag effective area and background rates\footnote{See e.g. \href{http://eastrogam.iaps.inaf.it/ scientific\_instrument.html}{\ag website}}. We have analyzed the simulated \ag spectrum combined with the \flat spectrum using the same approach as described in Section \ref{sec:data_analysis} to estimate the sensitivity reach of \ag.

CTA~\cite{cta_concept,cta_science} is the next-generation ground-based \gr\ observatory which will provide coverage of the energy band 0.03 -- 100~TeV. Assuming that the spectrum of NGC 1275 measured by MAGIC~\citet{magic_ngc1275} continues as a powerlaw toward higher energies, we consider the signal which can be observed with CTA. We consider the spectrum with  30 energy bins of equal width in log scale and assume that the statistics of the signal below 1 TeV is high enough so that the signal is dominated by 10\% systematic error. The sensitivity reach of CTA estimated with such an assumption is shown by the cyan dashed line in Fig. \ref{fig:exclusions_all}.

The resulting sensitivity of joint \ag+\flat +CTA observations is shown with dashed cyan line in Fig.~\ref{fig:exclusions_all} (right panel). As expected, the possibility to probe $E_{cr}$ values down to $\sim 0.1$~MeV results in the extension of the sensitivity range to the masses $m_a\sim 10^{-11}$~eV for the same level of the coupling $g$. Extension of the data to TeV energies expected with CTA allow to extend the exclusion region to $m_a\sim 10^{-7}$~eV. Such extension is particularly important since it will probe a part of the ALP dark matter parameter space. It will also completely cover the range of $m_a, g$ parameters in which the ALP-photon conversion could increase the transparency of the Universe for very-high-energy \gr s \cite{lower_limits}.

The ALP search method based on photon-ALP conversion in the Perseus cluster is complementary to another  method which will be used by CTA: measurement of the increase of transparency of the Universe to $\gamma$-rays from high-z Bl Lacs \cite{axion_cta}. Fig. \ref{fig:exclusions_all} (right panel) shows a comparison of the sensitivities of the two methods. One could see that the sensitivity of the search using particular source, NGC 1275 in Perseus cluster, is superior to that of the search based on distant BL Lacs, because of the proximity of NGC 1275 / Perseus cluster pair and because of the better knowledge of magnetic field properties in the cluster. Nevertheless, better constraints on galaxy cluster / galaxy group environments of \gr\ blazars and radio galaxies might result in a significant improvement of the sensitivity, extending the analysis presented in this paper to other nearby radio galaxies and blazars.

\section*{Acknowledgements} 

The authors acknowledge support by the state of Baden-W\"urttemberg through bwHPC. This work was supported by the Carl-Zeiss Stiftung through the grant ``Hochsensitive Nachweistechnik zur Erforschung des unsichtbaren Universums'' to the Kepler Center f{\"u}r Astro- und Teilchenphysik at the University of T{\"u}bingen.

\def\aj{AJ}%
\def\actaa{Acta Astron.}%
\def\araa{ARA\&A}%
\def\apj{ApJ}%
\def\apjl{ApJ}%
\def\apjs{ApJS}%
\def\ao{Appl.~Opt.}%
\def\apss{Ap\&SS}%
\def\aap{A\&A}%
\def\aapr{A\&A~Rev.}%
\def\aaps{A\&AS}%
\def\azh{AZh}%
\def\baas{BAAS}%
\def\bac{Bull. astr. Inst. Czechosl.}%
\def\caa{Chinese Astron. Astrophys.}%
\def\cjaa{Chinese J. Astron. Astrophys.}%
\def\icarus{Icarus}%
\def\jcap{J. Cosmology Astropart. Phys.}%
\def\jrasc{JRASC}%
\def\mnras{MNRAS}%
\def\memras{MmRAS}%
\def\na{New A}%
\def\nar{New A Rev.}%
\def\pasa{PASA}%
\def\pra{Phys.~Rev.~A}%
\def\prb{Phys.~Rev.~B}%
\def\prc{Phys.~Rev.~C}%
\def\prd{Phys.~Rev.~D}%
\def\pre{Phys.~Rev.~E}%
\def\prl{Phys.~Rev.~Lett.}%
\def\pasp{PASP}%
\def\pasj{PASJ}%
\def\qjras{QJRAS}%
\def\rmxaa{Rev. Mexicana Astron. Astrofis.}%
\def\skytel{S\&T}%
\def\solphys{Sol.~Phys.}%
\def\sovast{Soviet~Ast.}%
\def\ssr{Space~Sci.~Rev.}%
\def\zap{ZAp}%
\def\nat{Nature}%
\def\iaucirc{IAU~Circ.}%
\def\aplett{Astrophys.~Lett.}%
\def\apspr{Astrophys.~Space~Phys.~Res.}%
\def\bain{Bull.~Astron.~Inst.~Netherlands}%
\def\fcp{Fund.~Cosmic~Phys.}%
\def\gca{Geochim.~Cosmochim.~Acta}%
\def\grl{Geophys.~Res.~Lett.}%
\def\jcp{J.~Chem.~Phys.}%
\def\jgr{J.~Geophys.~Res.}%
\def\jqsrt{J.~Quant.~Spec.~Radiat.~Transf.}%
\def\memsai{Mem.~Soc.~Astron.~Italiana}%
\def\nphysa{Nucl.~Phys.~A}%
\def\physrep{Phys.~Rep.}%
\def\physscr{Phys.~Scr}%
\def\planss{Planet.~Space~Sci.}%
\def\procspie{Proc.~SPIE}%
\let\astap=\aap
\let\apjlett=\apjl
\let\apjsupp=\apjs
\let\applopt=\ao


\begin{thebibliography}{39}
\expandafter\ifx\csname natexlab\endcsname\relax\def\natexlab#1{#1}\fi

\bibitem[{{Abbott} \& {Sikivie}(1983)}]{abbott83}
{Abbott}, L.~F. \& {Sikivie}, P. 1983, Physics Letters B, 120, 133

\bibitem[{{Abramowski} {et~al.}(2013){Abramowski}, {Acero}, {Aharonian}, {Ait
  Benkhali}, {Akhperjanian}, {Ang{\"u}ner}, {Anton}, {Balenderan}, {Balzer},
  {Barnacka}, \& et~al.}]{hess_pks2155}
{Abramowski}, A., {Acero}, F., {Aharonian}, F., {et~al.} 2013, \prd, 88, 102003

\bibitem[{{Acharya} {et~al.}(2013){Acharya}, {Actis}, {Aghajani}, {Agnetta},
  {Aguilar}, {Aharonian}, {Ajello}, {Akhperjanian}, {Alcubierre},
  {Aleksi{\'c}}, \& et~al.}]{cta_concept}
{Acharya}, B.~S., {Actis}, M., {Aghajani}, T., {et~al.} 2013, Astroparticle
  Physics, 43, 3

\bibitem[{{Ahnen} {et~al.}(2016){Ahnen}, {Ansoldi}, {Antonelli}, {Antoranz},
  {Babic}, {Banerjee}, {Bangale}, {Barres de Almeida}, {Barrio}, {Becerra
  Gonz{\'a}lez}, {Bednarek}, {Bernardini}, {Biasuzzi}, {Biland}, {Blanch},
  {Bonnefoy}, {Bonnoli}, {Borracci}, {Bretz}, {Buson}, {Carmona}, {Carosi},
  {Chatterjee}, {Clavero}, {Colin}, {Colombo}, {Contreras}, {Cortina},
  {Covino}, {Da Vela}, {Dazzi}, {De Angelis}, {De Lotto}, {de O{\~n}a
  Wilhelmi}, {Delgado Mendez}, {Di Pierro}, {Dom{\'{\i}}nguez}, {Dominis
  Prester}, {Dorner}, {Doro}, {Einecke}, {Eisenacher Glawion}, {Elsaesser},
  {Fern{\'a}ndez-Barral}, {Fidalgo}, {Fonseca}, {Font}, {Frantzen}, {Fruck},
  {Galindo}, {Garc{\'{\i}}a L{\'o}pez}, {Garczarczyk}, {Garrido Terrats},
  {Gaug}, {Giammaria}, {Godinovi{\'c}}, {Gonz{\'a}lez Mu{\~n}oz}, {Gora},
  {Guberman}, {Hadasch}, {Hahn}, {Hanabata}, {Hayashida}, {Herrera}, {Hose},
  {Hrupec}, {Hughes}, {Idec}, {Kodani}, {Konno}, {Kubo}, {Kushida}, {La
  Barbera}, {Lelas}, {Lindfors}, {Lombardi}, {Longo}, {L{\'o}pez},
  {L{\'o}pez-Coto}, {Lorenz}, {Majumdar}, {Makariev}, {Mallot}, {Maneva},
  {Manganaro}, {Mannheim}, {Maraschi}, {Marcote}, {Mariotti},
  {Mart{\'{\i}}nez}, {Mazin}, {Menzel}, {Miranda}, {Mirzoyan}, {Moralejo},
  {Moretti}, {Nakajima}, {Neustroev}, {Niedzwiecki}, {Nievas Rosillo},
  {Nilsson}, {Nishijima}, {Noda}, {Orito}, {Overkemping}, {Paiano}, {Palacio},
  {Palatiello}, {Paneque}, {Paoletti}, {Paredes}, {Paredes-Fortuny},
  {Pedaletti}, {Persic}, {Poutanen}, {Prada Moroni}, {Prandini}, {Puljak},
  {Rhode}, {Rib{\'o}}, {Rico}, {Rodriguez Garcia}, {Saito}, {Satalecka},
  {Schultz}, {Schweizer}, {Sillanp{\"a}{\"a}}, {Sitarek}, {Snidaric},
  {Sobczynska}, {Stamerra}, {Steinbring}, {Strzys}, {Takalo}, {Takami},
  {Tavecchio}, {Temnikov}, {Terzi{\'c}}, {Tescaro}, {Teshima}, {Thaele},
  {Torres}, {Toyama}, {Treves}, {Vazquez Acosta}, {Verguilov}, {Vovk}, {Ward},
  {Will}, {Wu}, {Zanin}, {Pfrommer}, {Pinzke}, \& {Zandanel}}]{magic_ngc1275}
{Ahnen}, M.~L., {Ansoldi}, S., {Antonelli}, L.~A., {et~al.} 2016, \aap, 589,
  A33

\bibitem[{{Ajello} {et~al.}(2016){Ajello}, {Albert}, {Anderson}, {Baldini},
  {Barbiellini}, {Bastieri}, {Bellazzini}, {Bissaldi}, {Blandford}, {Bloom},
  {Bonino}, {Bottacini}, {Bregeon}, {Bruel}, {Buehler}, {Caliandro}, {Cameron},
  {Caragiulo}, {Caraveo}, {Cecchi}, {Chekhtman}, {Ciprini}, {Cohen-Tanugi},
  {Conrad}, {Costanza}, {D'Ammando}, {de Angelis}, {de Palma}, {Desiante}, {Di
  Mauro}, {Di Venere}, {Dom{\'{\i}}nguez}, {Drell}, {Favuzzi}, {Focke},
  {Franckowiak}, {Fukazawa}, {Funk}, {Fusco}, {Gargano}, {Gasparrini},
  {Giglietto}, {Glanzman}, {Godfrey}, {Guiriec}, {Horan}, {J{\'o}hannesson},
  {Katsuragawa}, {Kensei}, {Kuss}, {Larsson}, {Latronico}, {Li}, {Li}, {Longo},
  {Loparco}, {Lubrano}, {Madejski}, {Maldera}, {Manfreda}, {Mayer},
  {Mazziotta}, {Meyer}, {Michelson}, {Mirabal}, {Mizuno}, {Monzani},
  {Morselli}, {Moskalenko}, {Murgia}, {Negro}, {Nuss}, {Okada}, {Orlando},
  {Ormes}, {Paneque}, {Perkins}, {Pesce-Rollins}, {Piron}, {Pivato}, {Porter},
  {Rain{\`o}}, {Rando}, {Razzano}, {Reimer}, {S{\'a}nchez-Conde}, {Sgr{\`o}},
  {Simone}, {Siskind}, {Spada}, {Spandre}, {Spinelli}, {Takahashi}, {Thayer},
  {Torres}, {Tosti}, {Troja}, {Uchiyama}, {Wood}, {Wood}, {Zaharijas},
  {Zimmer}, \& {Fermi-LAT Collaboration}}]{fermi_ngc1275}
{Ajello}, M., {Albert}, A., {Anderson}, B., {et~al.} 2016, Physical Review
  Letters, 116, 161101

\bibitem[{{Aleksi{\'c}} {et~al.}(2010){Aleksi{\'c}}, {Antonelli}, {Antoranz},
  {Backes}, {Barrio}, {Bastieri}, {Becerra Gonz{\'a}lez}, {Bednarek},
  {Berdyugin}, {Berger}, {Bernardini}, {Biland}, {Blanch}, {Bock}, {Boller},
  {Bonnoli}, {Bordas}, {Borla Tridon}, {Bosch-Ramon}, {Bose}, {Braun}, {Bretz},
  {Camara}, {Ca{\~n}ellas}, {Carmona}, {Carosi}, {Colin}, {Colombo},
  {Contreras}, {Cortina}, {Cossio}, {Covino}, {Dazzi}, {De Angelis}, {De Cea
  del Pozo}, {De Lotto}, {De Maria}, {De Sabata}, {Delgado Mendez}, {Diago
  Ortega}, {Doert}, {Dom{\'{\i}}nguez}, {Dominis Prester}, {Dorner}, {Doro},
  {Elsaesser}, {Errando}, {Ferenc}, {Fonseca}, {Font}, {Garc{\'{\i}}a
  L{\'o}pez}, {Garczarczyk}, {Giavitto}, {Godinovi{\'c}}, {Hadasch}, {Herrero},
  {Hildebrand}, {H{\"o}hne-M{\"o}nch}, {Hose}, {Hrupec}, {Jogler}, {Klepser},
  {Kr{\"a}henb{\"u}hl}, {Kranich}, {Krause}, {La Barbera}, {Leonardo},
  {Lindfors}, {Lombardi}, {Longo}, {L{\'o}pez}, {Lorenz}, {Majumdar},
  {Makariev}, {Maneva}, {Mankuzhiyil}, {Mannheim}, {Maraschi}, {Mariotti},
  {Mart{\'{\i}}nez}, {Mazin}, {Meucci}, {Miranda}, {Mirzoyan}, {Miyamoto},
  {Mold{\'o}n}, {Moralejo}, {Nieto}, {Nilsson}, {Orito}, {Oya}, {Paoletti},
  {Paredes}, {Partini}, {Pasanen}, {Pauss}, {Pegna}, {Perez-Torres}, {Persic},
  {Peruzzo}, {Pochon}, {Prada}, {Prada Moroni}, {Prandini}, {Puchades},
  {Puljak}, {Reichardt}, {Reinthal}, {Rhode}, {Rib{\'o}}, {Rico},
  {R{\"u}gamer}, {Saggion}, {Saito}, {Saito}, {Salvati}, {S{\'a}nchez-Conde},
  {Satalecka}, {Scalzotto}, {Scapin}, {Schultz}, {Schweizer}, {Shayduk},
  {Shore}, {Sierpowska-Bartosik}, {Sillanp{\"a}{\"a}}, {Sitarek}, {Sobczynska},
  {Spanier}, {Spiro}, {Stamerra}, {Steinke}, {Storz}, {Strah}, {Struebig},
  {Suric}, {Takalo}, {Tavecchio}, {Temnikov}, {Terzi{\'c}}, {Tescaro},
  {Teshima}, {Torres}, {Vankov}, {Wagner}, {Weitzel}, {Zabalza}, {Zandanel},
  {Zanin}, {Neronov}, {Pfrommer}, {Pinzke}, {Semikoz}, \& {MAGIC
  Collaboration}}]{IC310_MAGIC}
{Aleksi{\'c}}, J., {Antonelli}, L.~A., {Antoranz}, P., {et~al.} 2010, \apjl,
  723, L207

\bibitem[{{Anastassopoulos} {et~al.}(2017){Anastassopoulos}, {Aune}, {Barth},
  {Belov}, {Br{\"a}uninger}, {Cantatore}, {Carmona}, {Castel}, {Cetin},
  {Christensen}, {Collar}, {Dafni}, {Davenport}, {Decker}, {Dermenev}, {Desch},
  {Eleftheriadis}, {Fanourakis}, {Ferrer-Ribas}, {Fischer}, {Garc{\'{\i}}a},
  {Gardikiotis}, {Garza}, {Gazis}, {Geralis}, {Giomataris}, {Gninenko},
  {Hailey}, {Hasinoff}, {Hoffmann}, {Iguaz}, {Irastorza}, {Jakobsen}, {Jacoby},
  {Jakov{\v c}i{\'c}}, {Kaminski}, {Karuza}, {Kralj}, {Kr{\v c}mar},
  {Kostoglou}, {Krieger}, {Laki{\'c}}, {Laurent}, {Liolios}, {Ljubi{\v
  c}i{\'c}}, {Luz{\'o}n}, {Maroudas}, {Miceli}, {Neff}, {Ortega},
  {Papaevangelou}, {Paraschou}, {Pivovaroff}, {Raffelt}, {Rosu}, {Ruz},
  {Ch{\'o}liz}, {Savvidis}, {Schmidt}, {Semertzidis}, {Solanki}, {Stewart},
  {Vafeiadis}, {Vogel}, {Yildiz}, \& {Zioutas}}]{new_cast_limits}
{Anastassopoulos}, V., {Aune}, S., {Barth}, K., {et~al.} 2017, Nature Physics,
  13, 584

\bibitem[{{Arias} {et~al.}(2012){Arias}, {Cadamuro}, {Goodsell}, {Jaeckel},
  {Redondo}, \& {Ringwald}}]{ringwald}
{Arias}, P., {Cadamuro}, D., {Goodsell}, M., {et~al.} 2012, \jcap, 6, 013

\bibitem[{{Asztalos} {et~al.}(2011){Asztalos}, {Carosi}, {Hagmann}, {Kinion},
  {van Bibber}, {Hotz}, {Rosenberg}, {Rybka}, {Wagner}, {Hoskins}, {Martin},
  {Sullivan}, {Tanner}, {Bradley}, \& {Clarke}}]{admx_ex}
{Asztalos}, S.~J., {Carosi}, G., {Hagmann}, C., {et~al.} 2011, Nuclear
  Instruments and Methods in Physics Research A, 656, 39

\bibitem[{{B{\"a}hre} {et~al.}(2013){B{\"a}hre}, {D{\"o}brich},
  {Dreyling-Eschweiler}, {Ghazaryan}, {Hodajerdi}, {Horns}, {Januschek},
  {Knabbe}, {Lindner}, {Notz}, {Ringwald}, {von Seggern}, {Stromhagen},
  {Trines}, \& {Willke}}]{alps2_ex}
{B{\"a}hre}, R., {D{\"o}brich}, B., {Dreyling-Eschweiler}, J., {et~al.} 2013,
  Journal of Instrumentation, 8, T09001

\bibitem[{{Berg} {et~al.}(2017){Berg}, {Conlon}, {Day}, {Jennings},
  {Krippendorf}, {Powell}, \& {Rummel}}]{ngc1275_xray_limits}
{Berg}, M., {Conlon}, J.~P., {Day}, F., {et~al.} 2017, \apj, 847, 101

\bibitem[{{Chelouche} {et~al.}(2009){Chelouche}, {Rabad{\'a}n}, {Pavlov}, \&
  {Castej{\'o}n}}]{chelouche08}
{Chelouche}, D., {Rabad{\'a}n}, R., {Pavlov}, S.~S., \& {Castej{\'o}n}, F.
  2009, \apjs, 180, 1

\bibitem[{{Cherenkov Telescope Array Consortium} {et~al.}(2017){Cherenkov
  Telescope Array Consortium}, {:}, {Acharya}, {Agudo}, {Samarai}, {Alfaro},
  {Alfaro}, {Alispach}, {Alves Batista}, {Amans}, \& et~al.}]{cta_science}
{Cherenkov Telescope Array Consortium}, T., {:}, {Acharya}, B.~S., {et~al.}
  2017, ArXiv e-prints [\eprint[arXiv]{1709.07997}]

\bibitem[{{Churazov} {et~al.}(2003){Churazov}, {Forman}, {Jones}, \&
  {B{\"o}hringer}}]{churazov}
{Churazov}, E., {Forman}, W., {Jones}, C., \& {B{\"o}hringer}, H. 2003, \apj,
  590, 225

\bibitem[{{De Angelis} {et~al.}(2017{\natexlab{a}}){De Angelis}, {Tatischeff},
  {Grenier}, {McEnery}, {Mallamaci}, {Tavani}, {Oberlack}, {Hanlon}, {Walter},
  {Argan}, \& et~al.}]{ag_science}
{De Angelis}, A., {Tatischeff}, V., {Grenier}, I.~A., {et~al.}
  2017{\natexlab{a}}, ArXiv e-prints [\eprint[arXiv]{1711.01265}]

\bibitem[{{De Angelis} {et~al.}(2017{\natexlab{b}}){De Angelis}, {Tatischeff},
  {Tavani}, {Oberlack}, {Grenier}, {Hanlon}, {Walter}, {Argan}, {von Ballmoos},
  {Bulgarelli}, {Donnarumma}, {Hernanz}, {Kuvvetli}, {Pearce}, {Zdziarski},
  {Aboudan}, {Ajello}, {Ambrosi}, {Bernard}, {Bernardini}, {Bonvicini},
  {Brogna}, {Branchesi}, {Budtz-Jorgensen}, {Bykov}, {Campana}, {Cardillo},
  {Coppi}, {De Martino}, {Diehl}, {Doro}, {Fioretti}, {Funk}, {Ghisellini},
  {Grove}, {Hamadache}, {Hartmann}, {Hayashida}, {Isern}, {Kanbach}, {Kiener},
  {Kn{\"o}dlseder}, {Labanti}, {Laurent}, {Limousin}, {Longo}, {Mannheim},
  {Marisaldi}, {Martinez}, {Mazziotta}, {McEnery}, {Mereghetti}, {Minervini},
  {Moiseev}, {Morselli}, {Nakazawa}, {Orleanski}, {Paredes}, {Patricelli},
  {Peyr{\'e}}, {Piano}, {Pohl}, {Ramarijaona}, {Rando}, {Reichardt},
  {Roncadelli}, {Silva}, {Tavecchio}, {Thompson}, {Turolla}, {Ulyanov},
  {Vacchi}, {Wu}, \& {Zoglauer}}]{ag_description}
{De Angelis}, A., {Tatischeff}, V., {Tavani}, M., {et~al.} 2017{\natexlab{b}},
  Experimental Astronomy, 44, 25

\bibitem[{{Duffy} \& {van Bibber}(2009)}]{duffy09}
{Duffy}, L.~D. \& {van Bibber}, K. 2009, New Journal of Physics, 11, 105008

\bibitem[{{Giannotti} {et~al.}(2016){Giannotti}, {Ruz}, {Vogel}, \& {IAXO
  Collaboration}}]{iaxo_limits}
{Giannotti}, M., {Ruz}, J., {Vogel}, J., \& {IAXO Collaboration}. 2016, in
  Proceedings of the 38th International Conference on High Energy Physics
  (ICHEP2016). 3-10 August 2016. Chicago, USA. Online at
  \href{http://pos.sissa.it/cgi-bin/reader/conf.cgi?confid=282}{http://pos.sissa.it/cgi-bin/reader/conf.cgi?confid=282},
  id.195, 195

\bibitem[{{Graham} {et~al.}(2015){Graham}, {Irastorza}, {Lamoreaux}, {Lindner},
  \& {van Bibber}}]{alps2_limits}
{Graham}, P.~W., {Irastorza}, I.~G., {Lamoreaux}, S.~K., {Lindner}, A., \& {van
  Bibber}, K.~A. 2015, Annual Review of Nuclear and Particle Science, 65, 485

\bibitem[{{Hochmuth} \& {Sigl}(2007)}]{hochmuth07}
{Hochmuth}, K.~A. \& {Sigl}, G. 2007, \prd, 76, 123011

\bibitem[{{Kuster} {et~al.}(2007){Kuster}, {Br{\"a}uninger}, {Cebri{\'a}n},
  {Davenport}, {Eleftheriadis}, {Englhauser}, {Fischer}, {Franz}, {Friedrich},
  {Hartmann}, {Heinsius}, {Hoffmann}, {Hoffmeister}, {Joux}, {Kang},
  {K{\"o}nigsmann}, {Kotthaus}, {Papaevangelou}, {Lasseur}, {Lippitsch},
  {Lutz}, {Morales}, {Rodr{\'{\i}}guez}, {Str{\"u}der}, {Vogel}, \&
  {Zioutas}}]{cast_ex}
{Kuster}, M., {Br{\"a}uninger}, H., {Cebri{\'a}n}, S., {et~al.} 2007, New
  Journal of Physics, 9, 169

\bibitem[{{Majorovits} \& {Javier Redondo for the MADMAX Working
  Group}(2016)}]{madmax_ex}
{Majorovits}, B. \& {Javier Redondo for the MADMAX Working Group}. 2016, ArXiv
  e-prints [\eprint[arXiv]{1611.04549}]

\bibitem[{{Marsh}(2017)}]{marsh17}
{Marsh}, D.~J.~E. 2017, ArXiv e-prints [\eprint[arXiv]{1712.03018}]

\bibitem[{{Marsh} {et~al.}(2017){Marsh}, {Russell}, {Fabian}, {McNamara},
  {Nulsen}, \& {Reynolds}}]{m87_xray_limits}
{Marsh}, M.~C.~D., {Russell}, H.~R., {Fabian}, A.~C., {et~al.} 2017, \jcap, 12,
  036

\bibitem[{{Meyer} \& {Conrad}(2014)}]{axion_cta}
{Meyer}, M. \& {Conrad}, J. 2014, \jcap, 12, 016

\bibitem[{{Meyer} {et~al.}(2013){Meyer}, {Horns}, \& {Raue}}]{lower_limits}
{Meyer}, M., {Horns}, D., \& {Raue}, M. 2013, \prd, 87, 035027

\bibitem[{{Mirizzi} \& {Montanino}(2009)}]{mirizzi09}
{Mirizzi}, A. \& {Montanino}, D. 2009, \jcap, 12, 004

\bibitem[{{Neronov} {et~al.}(2010){Neronov}, {Semikoz}, \& {Vovk}}]{ic310}
{Neronov}, A., {Semikoz}, D., \& {Vovk}, I. 2010, \aap, 519, L6

\bibitem[{{Payez} {et~al.}(2015){Payez}, {Evoli}, {Fischer}, {Giannotti},
  {Mirizzi}, \& {Ringwald}}]{sn1987_limits}
{Payez}, A., {Evoli}, C., {Fischer}, T., {et~al.} 2015, \jcap, 2, 006

\bibitem[{{Peccei} \& {Quinn}(1977)}]{peccei}
{Peccei}, R.~D. \& {Quinn}, H.~R. 1977, Physical Review Letters, 38, 1440

\bibitem[{{Preskill} {et~al.}(1983){Preskill}, {Wise}, \&
  {Wilczek}}]{preskill83}
{Preskill}, J., {Wise}, M.~B., \& {Wilczek}, F. 1983, Physics Letters B, 120,
  127

\bibitem[{Primakoff(1951)}]{primakoff}
Primakoff, H. 1951, Phys. Rev., 81, 899

\bibitem[{{Raffelt}(1996)}]{raffelt96}
{Raffelt}, G.~G. 1996, {Stars as laboratories for fundamental physics : the
  astrophysics of neutrinos, axions, and other weakly interacting particles}

\bibitem[{Ressell(1991)}]{ressell91}
Ressell, M.~T. 1991, Phys. Rev. D, 44, 3001

\bibitem[{{Ringwald}(2014)}]{ringwald14}
{Ringwald}, A. 2014, in Journal of Physics Conference Series, Vol. 485, Journal
  of Physics Conference Series, 012013

\bibitem[{{Taylor} {et~al.}(2006){Taylor}, {Gugliucci}, {Fabian}, {Sanders},
  {Gentile}, \& {Allen}}]{magnetic_field}
{Taylor}, G.~B., {Gugliucci}, N.~E., {Fabian}, A.~C., {et~al.} 2006, \mnras,
  368, 1500

\bibitem[{{Vogel} {et~al.}(2013){Vogel}, {Avignone}, {Cantatore}, {Carmona},
  {Caspi}, {Cetin}, {Christensen}, {Dael}, {Dafni}, {Davenport}, {Derbin},
  {Desch}, {Diago}, {Dudarev}, {Eleftheriadis}, {Fanourakis}, {Ferrer-Ribas},
  {Galan}, {Garcia}, {Garza}, {Geralis}, {Gimeno}, {Giomataris}, {Gninenko},
  {Gomez}, {Hailey}, {Hiramatsu}, {Hoffmann}, {Iguaz}, {Irastorza}, {Isern},
  {Jaeckel}, {Jakovcic}, {Kaminski}, {Kawasaki}, {Krcmar}, {Krieger}, {Lakic},
  {Lindner}, {Liolios}, {Luzon}, {Ortega}, {Papaevangelou}, {Pivovaroff},
  {Raffelt}, {Redondo}, {Ringwald}, {Russenschuck}, {Ruz}, {Saikawa},
  {Savvidis}, {Sekiguchi}, {Shilon}, {Silva}, {ten Kate}, {Tomas}, {Troitsky},
  {van Bibber}, {Vedrine}, {Villar}, {Walckiers}, {Wester}, {Yildiz}, \&
  {Zioutas}}]{iaxo_ex}
{Vogel}, J.~K., {Avignone}, F.~T., {Cantatore}, G., {et~al.} 2013, ArXiv
  e-prints [\eprint[arXiv]{1302.3273}]

\bibitem[{{Weinberg}(1978)}]{weinberg78}
{Weinberg}, S. 1978, Physical Review Letters, 40, 223

\bibitem[{{Wilczek}(1978)}]{wilczek78}
{Wilczek}, F. 1978, Physical Review Letters, 40, 279

\end{thebibliography}

\end{document}